\documentclass[aps,amsmath,amssymb,reprint,]{revtex4-1}
\usepackage{graphicx}
\usepackage{dcolumn}
\usepackage{bm}
\usepackage[utf8]{inputenc}
\usepackage[T1]{fontenc}
\usepackage{mathptmx}
\usepackage{etoolbox}
\makeatletter
\def\@email#1#2{%
 \endgroup
 \patchcmd{\titleblock@produce}
  {\frontmatter@RRAPformat}
  {\frontmatter@RRAPformat{\produce@RRAP{*#1\href{mailto:#2}{#2}}}\frontmatter@RRAPformat}
  {}{}
}
\makeatother

\usepackage{hyperref}
\hypersetup{colorlinks,linkcolor={blue},citecolor={blue},urlcolor={blue}}
\begin{document}
\preprint{AIP/123-QED}

\title {Geometric Brownian Information Engine: Essentials for the best performance }

\author{Rafna Rafeek}
\affiliation{Department of Chemistry and Center for Molecular and Optical Sciences \& Technologies, Indian Institute of Technology Tirupati, Yerpedu 517619, Andhra Pradesh, India}
\author{Syed Yunus Ali}
\affiliation{Department of Chemistry and Center for Molecular and Optical Sciences \& Technologies, Indian Institute of Technology Tirupati, Yerpedu 517619, Andhra Pradesh, India}
\author{Debasish Mondal}
 \email{debasish@iittp.ac.in}
\affiliation{Department of Chemistry and Center for Molecular and Optical Sciences \& Technologies, Indian Institute of Technology Tirupati, Yerpedu 517619, Andhra Pradesh, India}

\date{\today}% It is always \today, today,

\begin{abstract}
 We investigate a Geometric Brownian Information Engine (GBIE) in the presence of an error-free feedback controller that transforms the information gathered on the state of Brownian particles entrapped in monolobal geometric confinement into extractable work. Outcomes of the information engine depend on the reference measurement distance $x_m$, feedback site $x_f$ and the transverse force $G$.  We determine the benchmarks for utilizing the available information in an output work and the optimum operating requisites for best work extraction. Transverse bias force ($G$) tunes the entropic contribution in the effective potential and hence the standard deviation ($\sigma$) of the equilibrium marginal probability distribution. We recognize that the amount of extracted work reaches a global maximum when $x_f = 2x_m$ with $x_m \sim 0.6\sigma$, irrespective of the extent of the entropic limitation. Because of the higher loss of information during the relaxation process, the best achievable work of a GBIE is lower in an entropic system. The feedback regulation also bears the unidirectional passage of particles. The average displacement increases with growing entropic control and is maximum when $x_m \sim 0.81\sigma$. Finally, we explore the efficacy of the information engine, a quantity that regulates the efficiency in utilizing the information acquired. With $x_f=2x_m$, the maximum efficacy reduces with increasing entropic control and shows a cross over from $2$ to $11/9$. We discover that the condition for the best efficacy depends only on the confinement length scale along the feedback direction.  The broader marginal probability distribution accredits the increased average displacement in a cycle and the lower efficacy in an entropy-dominated system.
\end{abstract}

\maketitle

\begin{quotation}

\end{quotation}

\section{INTRODUCTION}
In 1867, Maxwell proposed a hypothetical experiment (Maxwell's demon) that inspects gas molecules in a single heat bath and utilizes the obtained information to extract work, thus apparently violating the second law of thermodynamics \cite{Rex2003maxwell,Rex2017maxwell}.  Resolving of the paradox unveiled the connection between the thermodynamic entropy and information gathered on measurement \cite{Landaueribm1961,Bennett1982intjtphys,Brillouin1951jcp}. The Szilard's engine that involves a feedback-controlled measurement process and work is extracted using the collected information, serves as a foremost benchmark for this apparent paradox \cite{Szilard1929zphys}. Recently, Sagawa and Ueda offered the quantitative association between the entropy and information in the Information-fluctuation theorem \cite{Sagwa2008prl,Sagawa2009prl,Sagawa2010prl}, which designates the bound on the work\cite{Jarzynski1997prl} obtained from the available information. These developments emanated the notion of information engines, a system that extracts work from a single heat bath using the mutual information earned by the measurement. Execution of a feedback protocol \cite{Sagwa2008prl,Sagawa2009prl,Sagawa2010prl,Horowitz2010pre,Abreu2011epl,Pal2014pre,Ashida2014pre,Kim2011prl,Bruschi2015pre,Goold2016jphysA,Lopez2008prl,Toyabe2010natphys,Berut2012nat,Park2016pre,Paneru2018pre,Paneru2018prl,Dago2021,Paneru2020natcommun,Paneru2020}, however not limited to \cite{Blickle2012natphys,Kumaripre2020,Holubecpre2020,Zakineentropy2017,GomezFP2021}, is a widely popular mechanism to devise an information engine. Due to their consequences in living systems, several variants of information engines at the mesoscopic level have been studied theoretically, both classical  \cite{Abreu2011epl,Pal2014pre,Ashida2014pre} and quantum systems \cite{Sagwa2008prl,Kim2011prl,Bruschi2015pre,Goold2016jphysA}, and validated through experiments \cite{Berut2012nat,Paneru2018prl,Paneru2018pre, Paneru2020natcommun,Paneru2020, Dago2021}. In many occurrences, it involves a Brownian particle as a functional substance \cite{Lopez2008prl,Toyabe2010natphys,Berut2012nat, Paneru2018prl,Paneru2018pre,Park2016pre}.\\

The Brownian information engines are commonly realized by confining the particle in monostable or bistable optical traps and implementing an appropriate feedback controller across an operating direction. The upper bound of the achievable work from a Brownian information engine and its optimum functional recipe have been explored recently \cite{Park2016pre,Paneru2018prl,Paneru2018pre}. The capacity of work extraction from such Brownian information engines depends on the strength (frequency) of the confining potential as the latter influences both measurement unpredictability and the relaxation process after the feedback operation. Therefore, the standard deviation of the equilibrium distribution of the particle inside the confining potential plays a central role in determining the best performance prescription. Scrutiny on the total information accumulated through the measurement process and the loss during the relaxation step shows that the Brownian Information Engine can act as a lossless engine under an error-free estimation \cite{Paneru2018prl}.\\

An overwhelming majority of these studies \cite{Lopez2008prl,Paneru2018prl,Paneru2018pre,Toyabe2010natphys,Martinez2016natphys,Park2016pre,Sahae2021pnas,Koski2014pnas,Chiuchi2018pra,Berut2012nat,Chiuchi2015,Proesmansprl2020,Proesmanspre2020} but \cite{Ali2021jcp} uses harmonic or bistable energetic confining potential to set up a Brownian information engine. Therefore, the following immediate interests emerge: (a) Can one actualize a Brownian information engine without external confining potentials? (b) If so, what will be the underlying working principle and performance ability? Recently, we have detailed one of such type; a Geometrical Brownian Information Engine \cite{Ali2021jcp}. We examined the motion of a free Brownian particle inside a 2-D narrow channel (mesoscopic scale) with varying width across the transport direction. By introducing an appropriate feedback controller, we have determined the upper bound of the extractable work. The particles confined in such geometry with uneven boundaries experiences an effective entropic potential along the transport direction. The entropic potential appears as a logarithmic function of the phase-space and is scaled with thermal energy. The equilibrium marginal probability distribution in reduced dimension shapes an inverted parabolic distribution in a purely energy-controlled regime. This affects both the total information gathered during the process and the amount of unavailable information caused by the relaxation process. Consequently, the upper bound of the maximum achievable work from a Brownian information engine in a purely entropy managed condition is less ($(5/3 -2\ln2)k_BT$) \cite{Ali2021jcp} than an analogous energetic engine ($k_BT/2$) \cite{Ashida2014pre}. Therefore, it will be interesting to explore the working policy of such GBIE and its outgrowths in detail and, hence, compare the  best performance requirements of an entropy-driven information engine with an analogous energetic device. Other than utilizing available information as useful work, the feedback process results in a unidirectional passage of the particle. Thus, analyzing how entropic limitation impacts the average displacement per cycle will also be crucial. The efficacy of an information engine is another observable of attention \cite{Sagawa2010prl}. The efficacy measures the uses of the information gathered through a measurement and gets influenced by relaxation pathways. The information loss during the relaxation in a GBIE differs from its energetic analogue, and it will be exciting to examine how the efficacy develops in increasing entropic control.\\

   In this context, it is noteworthy that the diffusive transport of micro-objects inside a narrow channel has received substantial attention in the recent past \cite{Zwanzig1992jpc,Reguera2001pre,Reguera2006prl,Burada2007pre,Burada2008prl,Burada2009epjb,Mondal2010jcp1,Mondal2010pre,Mondal2012pre,Quan2008jcp,Kalinay2005jcp,Mondal2010jcp2,Das2012jcp1,Das2012jcs,Malgaretti2019jcp,Burada2008biosyst,Arango2020jcp,Burada2009epl,Mondal2011jcp,Das2012pre, Das2014jcp}. Understanding such constrained motion is essential in biological processes like passage ions through the membrane \cite{Zhou2010jpcl}, translocation of polymers through narrow pores \cite{Muthukumar2003jcp,Mondal2016jcp1,Mondal2016jcp2} and chemical reactions in a constrained space \cite{Zhou1991jcp,Guerin2021communchem}.
In 1992, R. Zwanzig derived the theoretical formulation of diffusion inside a restrained channel with irregular boundaries \cite{Zwanzig1992jpc}. The diffusion process reduces into a one-dimensional Ficks-Jacob equation in which the effect of varying curvature is considered through an effective entropic potential of the form $k_BT\ln(\Omega(x))$, where $\Omega(x)$ represent the phase-space of the device. Interestingly, similar type of logarithmic potential appears as a working potential in various biophysical processes, such as optically trapped cold atoms \cite{Kessler2010prl,Kessler2012prl,Lutz2013nphys,Dechant2011prl}, DNA unzipping events \cite{Poland1966jcp1,Poland1966jcp2,Kaiser2014jpysa,Bar2007prl,Fogedby2007prl} and  many others \cite{Dyson1962jmp,Spohn1987jsp,Chavanis2002pre,Manning1969jcp,Levine2005jsp,Ray2020jcp}.
\\

\begin{figure}
    \centering
    \includegraphics[width=0.45\textwidth]{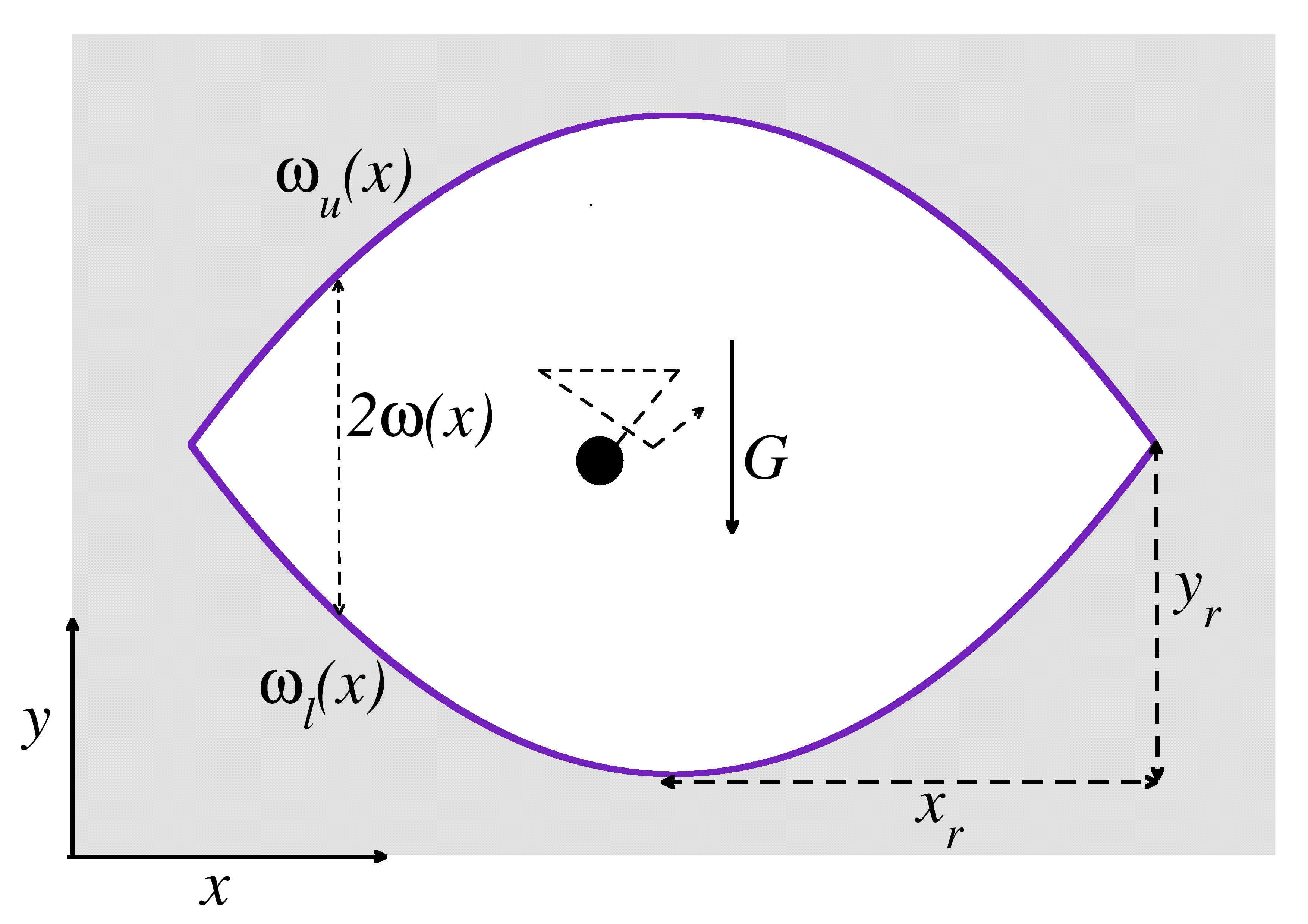}
    \caption{Schematic illustration of two-dimensional mono-lobal confinement. $\omega_u (x)$ and $\omega_l(x)$ are the boundary functions of the confinement. $x_r$ and $y_r$ are characteristic length scales that describe the confinement boundary. $\omega(x)$ is the local half-width at $x$. $G$ denotes the external transverse force acting orthogonal to the feedback direction.}
    \label{f1}
\end{figure}

This paper documents the working principles and provisions to the best production of a GBIE.  We study a Brownian particle trapped in a 2-dimensional monostable spatial confinement and restrained to a constant bias force ($G$) perpendicular to the longitudinal direction as shown in Fig.~\ref{f1} in the spirit of \cite{Ali2021jcp}. The transverse external force $G$ regulates the entropic contribution to the effective potential. We then introduce a feedback controller that consists of three steps: measurement, feedback and relaxation to complete the cycle, as depicted in Fig.~\ref{f2}. The particle inside the uneven confinement experiences an effective potential across the feedback direction. 
Because of the nontrivial interplay between the thermal fluctuations, phase space-dependent effective potential and the external force ($G$), the achievements of the information engines would alter significantly for varying measurement distances and feedback locations. We identified the favourable condition that the device acted as an engine and explored the optimum requisites to achieve maximum work. Using generalized integral-fluctuation relation, we have shown that a GBIE can transform all the available information to output work and, therefore, perform as a lossless information engine. We have also explored the influence of the entropic control on the other important outcomes of engine, such as the average movement per cycle and efficacy. We have compared our result with the performance ability of an energetic Brownian information engine and thus rendered a thorough understanding of the consequences of the entropic restriction.  

\section{MODEL AND METHOD}
\subsection{\label{app:subsec}Brownian particle in a Geometric confinement}
We consider a two-dimensional overdamped Brownian particle in a geometric confinement subjected to an external constant force $G$, acting along the transverse direction (as shown in Fig.~\ref{f1}).
Neglecting the inertial force, the dynamics of particle can be described by the following Langevin equation:
\begin{equation}\label{meq1}
\begin{aligned}
   \frac{d \vec{r}}{dt} &= -G\hat{e_y} +\vec{\zeta}(t), \\
\end{aligned}
\end{equation}
where the $\vec{r}$ denotes the position of the particle in two dimension, $\vec{r}=x \hat{e_x} + y \hat{e_y}$. $\hat{e_j}$ is the unit position vector along $j^{th}$-direction and $\vec{\zeta}(t)$ is the Gaussian white noise with following properties:
\begin{equation}\label{meq2} 
\begin{aligned}
\left \langle \zeta_j (t)\right \rangle &= 0, \; &for \; j =x,y \\
\left \langle \zeta_i (t) \zeta_j ({t}')\right \rangle &= 2D\delta_{ij}\delta (t-{t}'), \; & for\;i,j=x,y.
\end{aligned}
\end{equation}
Where $D=k_BT$ and $\langle ... \rangle$ denotes the averaged realisation. We have considered that the frictional coefficient of the particle is unity. The geometric confinement can be generated by imposing non-interactive static boundaries.
We describe the upper and lower walls, as depicted in the Fig.~\ref{f1}, using the boundary functions, $\omega_u(x) = -ax^2+c$ and $\omega_l(x) = -\omega_u(x)$, respectively. Where, $a$ and $c$ are constant confinement parameters. Therefore, the length scales along the $x$ and $y$-directions are $x_r$ ($=\sqrt{c/a}$) and $y_r$ ($=c$), respectively. The local half-width $\omega(x)$ measures the spatially varying cross-section of the confinement and  can be written as $\omega(x) = [\omega_u(x)-\omega_l(x)]/2$. The alternative Fokker-Planck description of the process (Eq.~\ref{meq1}-\ref{meq2}) can be expressed as: \cite{Burada2008prl,Burada2009epjb,Mondal2010pre,risken,gard,coxmiller} :
\begin{eqnarray}\label{meq5}
\frac{\partial}{ \partial t} p(x, y, t) & = & D\frac{\partial}{\partial x} \left \{  e^{\frac{-\psi(x,y)}{D}} \frac{\partial}{\partial x} e^{\frac{\psi(x,y)}{D}}p(x,y,t)\right \} \\ \nonumber & + & D\frac{\partial}{\partial y}\left \{ e^{\frac{-\psi(x,y)}{D}} \frac{\partial}{\partial y} e^{\frac{\psi(x,y)}{D}}p(x,y,t)\right \},
\end{eqnarray}
where,  $\psi(x,y) = Gy$ is a potential function and $p(x,y,t)$ is the probability distribution function of particle in $(x,y)$ at time $t$. %Since it is challenging to solve the two-dimensional Smoluchowski equation (Eq. ~\ref{meq5}) analytically, there is a need to reduce the dimensionality of the problem. 
%position $\vec{r}(x,y)$ at time $t$ as  
When the length scale along $x$-direction ($x_r$) is much larger than that along $y$-direction ($y_r$), one can assume a fast local equilibrium along the $y$-direction \cite{Zwanzig1992jpc,Reguera2001pre}. In this context, we define a position dependent potential function $A(x)$ as:

\begin{equation}\label{meq6}
    \begin{aligned}
  \exp\left(\frac{-A(x)}{D}\right) &= \int dy \exp\left(\frac{-\psi(x,y)}{D}\right).
    \end{aligned}
\end{equation}

If $\rho(y;x)$ is a conditional local equilibrium  distribution of $y$ for a given $x$ and $\rho(y;x)$ is normalised to unity  in  $y$, one can write:
\begin{equation}\label{meq6a}
    \begin{aligned}
  \rho(y;x)=\exp\left(\frac{A(x)}{D}\right)  \exp\left(\frac{-\psi(x,y)}{D}\right).
    \end{aligned}
\end{equation}

The fast-local equilibrium approximation along the transverse direction guides us to \cite{Zwanzig1992jpc,Reguera2001pre,Reguera2006prl,Burada2007pre, Burada2008prl}
\begin{equation}\label{meq6b}
    \begin{aligned}
  p(x,y,t)\simeq\rho(y;x)P(x,t).
    \end{aligned}
\end{equation}

Where, we express the marginal probability distribution function $P(x,t)$ as:
\begin{equation}\label{meq7}
    \begin{aligned}
    P(x,t) &= \int_{\omega_l(x)} ^{\omega_u(x)} p(x,y,t)dy.
    \end{aligned}
\end{equation}
\begin{figure}
    \centering
    \includegraphics[width=0.45\textwidth]{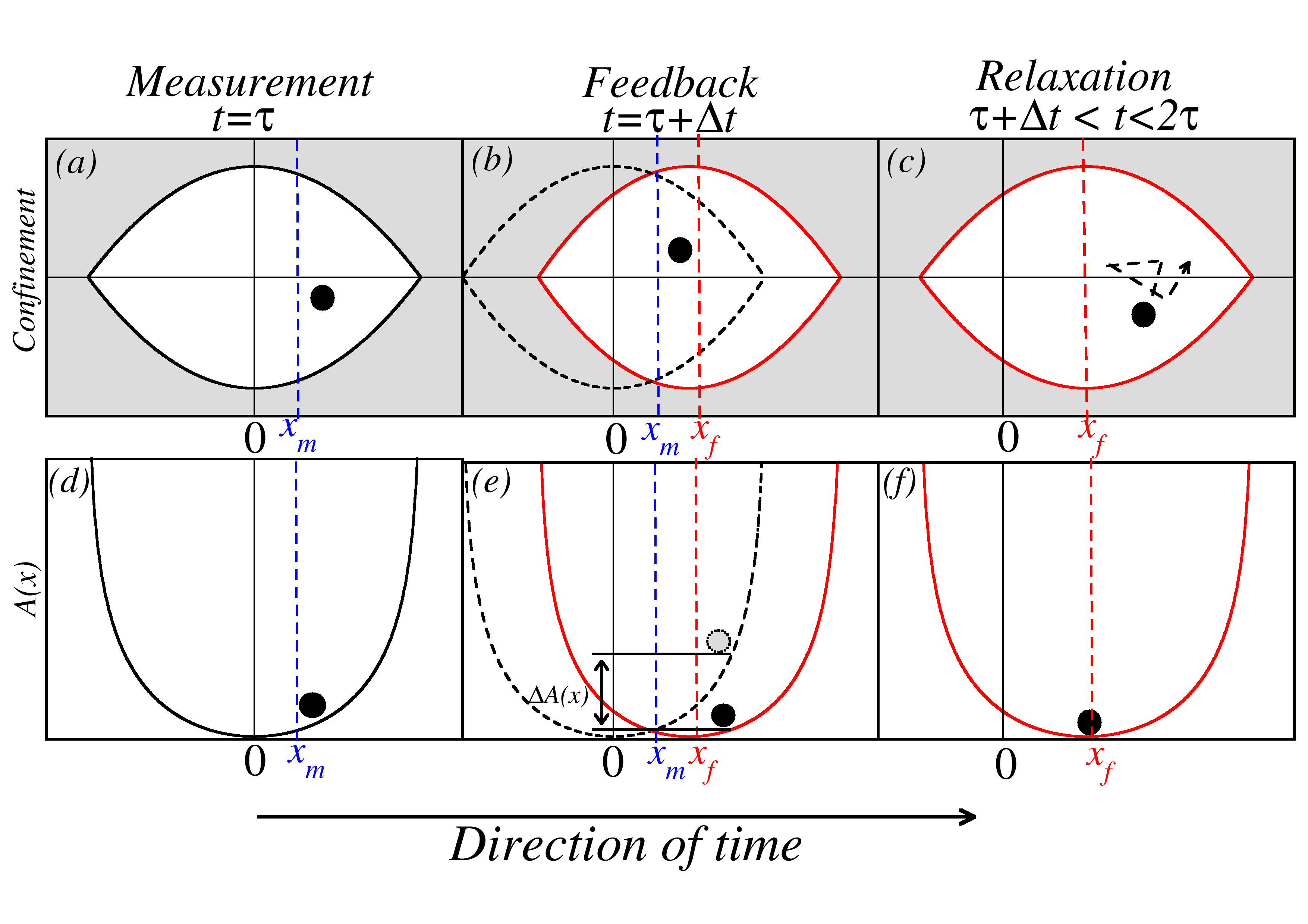}
     \caption{Schematic representation of a feedback protocol associated to a Geometric Brownian Information Engine (GBIE) during the time interval $\tau \leq t \leq 2\tau$. The feedback regulator consists of three steps: (a) Measurement: At $t=\tau$, the confinement centre is at zero ($\lambda(\tau) = 0$) and we measure the position ($x$) of the particle. We check whether $x\ge x_m$ or not, where $x_m$ is a measurement position (indicated by the vertical blue colored dashed line). (b) Feedback: If $x \ge x_m$, we shift the confinement center instantaneously to the feedback site $\lambda(\tau+\Delta t ) = x_f$ ( represented by red colored solid line). Otherwise, we keep the confinement centre unaltered $\lambda(\tau+\Delta t) = 0$. (c) Relaxation: Particle is allowed to relax then with the fixed confinement center at feedback site ($x_f$) until the next cycle begins (up to $t=2\tau$). (d-f) Illustration of the feedback regulation in terms of the effective geometric potential.
    }
    \label{f2}
\end{figure}

Using Eqs.~(\ref{meq6}-\ref{meq7}), the two-dimensional Smoluchowski Eq.~(\ref{meq5}) reduces into a Ficks-Jacobs equation in reduced dimension \cite{Zwanzig1992jpc,Reguera2001pre,Reguera2006prl,Burada2007pre,Burada2008biosyst,Burada2008prl,Burada2009epjb,Mondal2010pre,Mondal2010jcp1,Kalinay2005jcp,Mondal2012pre,Das2012jcp1,Das2012jcs,Mondal2010jcp2,Quan2008jcp,Mondal2011jcp,Burada2009epl,Das2012pre}:
\begin{equation}
    \begin{aligned}\label{meq8}
    \frac{\partial}{ \partial t} P(x,t) &= \frac{\partial}{\partial x}  \left \{ D  \frac{\partial}{\partial x} P(x,t) +  A'(x)P(x,t) \right\}, 
    \end{aligned}
\end{equation}
where, $A(x)$ is the effective potential experienced by the particle in reduced dimension:
\begin{equation}\label{meq9}
    \begin{aligned}
    A(x) = -D \ln \bigg[\frac{2D}{G} \sinh \bigg( \frac{G\omega(x)}{D} \bigg) \bigg].
    \end{aligned}
\end{equation}
Thus, the effective potential depends on the external transverse force $G$, the thermal energy $D$ and the geometry of the confinement in a non-trivial way. 
In the limit of $G/D \gg 1$, the effective potential reduces to $A(x)=-G\omega(x)$ and popularly denoted as energy-dominated situation. 
In the opposite limit $G/D \ll 1$, the effective potential becomes independent of $G$ with a logarithmic form, and the potential is purely entropic in nature;
\begin{equation}\label{meq10}
\begin{aligned}
   A(x) & = -G\omega(x), \;\;\;\; \text{for} \;\; \frac{G}{D} \gg 1, &\\
    &= -D \ln[2\omega(x)], \;\;\;\;\text{for} \;\; \frac{G}{D} \ll 1.
\end{aligned}
\end{equation}

\subsection{\label{app:subsec} GBIE: Feedback protocol}
We construct an information engine consisting of Brownian particles trapped in geometric confinement and subjected to a feedback control as illustrated in Fig.~\ref{f2}.
Each cycle consists of three steps: measurement, feedback and relaxation. As shown in the lower panel of Fig.~\ref{f2}, the particle, confined in a mono-lobal trap with uneven $\omega(x)$ along $x$ direction, experiences an effective potential $A(x-\lambda(t))$, where the $x$ is the position of the particle, $\lambda(t)$ is the centre of the confinement at time $t$. Initially, we take $\lambda(0)=0$. Once the thermal equilibrium is reached, i.e. at $t=\tau$, we perform a 'measurement' to determine the position of the particle $x$. We define a reference measurement distance at $x_m$. If the particle crosses the measurement distance ($x \geq x_m$), we shift the confinement centre instantaneously to $x_f$ ( i.e. $\lambda( \tau+\Delta t)=x_f$ and  $\Delta t \rightarrow 0$. In other words, the position of the effective potential centre also changes to $x_f$. Otherwise ($x < x_m$), we leave the confinement centre unaltered (i.e. $\lambda(\tau)=0$). In this scenario, we do not employ any feedback ($x_f=0$), and the centre of the effective potential remains unchanged. After the feedback, the particle relaxes with fixed $\lambda(\tau)$ until the next feedback. We set the time scale of the feedback protocol $\tau$ is much larger than the characteristic relaxation time scale of the system ($\tau \gg \tau_r$).  As the shift is instantaneous (error-free) and particles always return to the equilibrium state, the change in the potential energy can fully be converted into work.
Therefore, the work $-W(x)$ related to the measurement process can be written as:
\begin{equation}\label{feq1}
\begin{aligned}
   -W(x) & = A(x)-A(x-x_f), \;\;\;\; \text{if} \;\; x \geq x_m, &\\
    &= 0, \;\;\;\;\text{if} \;\; x < x_m.
\end{aligned}
\end{equation}
The process is repeated and the average extracted work is obtained as:
\begin{equation}\label{feq2}
    \begin{aligned}
       -\langle W \rangle = -\int_{-x_r}^{x_r} dxP_{eq}(x)W(x),
    \end{aligned}
\end{equation}
where $x_r=\sqrt{c/a}$ is the confinement length scale along $x$-axis and $P_{eq}(x) = lim_{t \rightarrow \infty}P(x,t)$ is the equilibrium marginal probability distribution. It is worthwhile to mention that the particle does not perform any work here of its own. Particles are transported due to the change in the potential centre during the feedback process along the
direction of transport. As the shift of the confinement
centre is instantaneous, there is no time for heat dissipation. Consequently, the change in potential energy is converted into extractable work. \\

 Next, we evaluate the net information acquired to examine the upper bound of the extracted work. The term {\it information} is related to the uncertainty of occurrence or surprisal of a certain event. Information related to an event $Y$ increases as the probability of the same ($P(Y)$) decreases. When $P(Y)$ tends to unity ($\sim 1$), the surprisal of the event is almost zero. On the other hand, if $P(Y)$ is extremely low  (close to zero), the surprisal of the event diverges. Therefore one can define the information related to an event $Y$ as $I(Y)=-ln(P(Y))$. In the present study, we consider an error-free feedback mechanism. In this situation, the net information grossed is equivalent to
the Shannon entropy of the particle at initial equilibrium since the Shannon entropy after the measurement is zero. Therefore for an error-free measurement process, the information can be expressed as \cite{Sagawa2010prl}, \cite{Ashida2014pre}, \cite{Ali2021jcp}:
\begin{equation}\label{feq3}
    \begin{aligned}
       \langle I \rangle = -\int_{-x_r}^{x_r} dxP_{eq}(x)ln[P_{eq}(x)].
    \end{aligned}
\end{equation}

 To estimate the unavailable information, we consider the reverse protocol: The particle is initially in equilibrium with the confinement location at $\lambda(t)=x_f$, and we shift the center back to $\lambda(t)=0$ suddenly irrespective of the position of the particle. For an error free (almost) measurement, the average unavailable information is \cite{Sagawa2010prl}, \cite{Ashida2014pre}, \cite{Ali2021jcp}:
\begin{equation}\label{feq4}
    \begin{aligned}
       \langle I_u \rangle = -\int_{-x_r}^{x_m} dxP_{eq}(x)ln[P_{eq}(x)] -\int_{x_m}^{x_r} dxP_{eq}(x)ln[P_{eq}(x-x_f)].
    \end{aligned}
\end{equation}
The unavailable information ($\langle I_u \rangle$) is an important quantity since it limits the possible work extraction. A higher $\langle I_u \rangle$ lowers the work extraction. We refer \cite{Ashida2014pre} for further details on total information and the unavailable information related to an event. Other physical observable of interest is the efficacy ($\gamma$) of the feedback control. The $\gamma$  measures how efficiently the device utilizes the net acquired information in the feedback protocol.  Using the concept of the generalised Jarzynski equality \cite{Jarzynski1997prl,Park2016pre,Paneru2018prl,Paneru2018pre}, $\gamma$ can be written as:
\begin{equation}\label{feq5}
    \begin{aligned}
        \gamma &= \langle \exp(-\beta W) \rangle  \\ &=\int_{-x_r}^{x_m} dxP_{eq}(x) +\int_{x_m}^{x_r} dxP_{eq}(x-x_f),
    \end{aligned}
\end{equation}
for an error-free measurement protocol. Finally, the average step ($\langle \Delta x \rangle$) per feedback cycle can be calculated as:
\begin{equation}\label{feq6}
    \begin{aligned}
        \langle \Delta x \rangle = x_f \int_{x_m}^{x_r}dxP_{eq}(x).
    \end{aligned}
\end{equation}

To proceed further, we discuss the ranges of the measurement ($x_m$) and feedback positions ($x_f$). In the present context, the measurement position can be set at any allowed position inside the confinement along the $x$-direction, $-x_r < x_m < x_r$.  Noticeably, the particle can never reach the terminal position $\pm x_{r}$ precisely. Associated uncertainty and hence the information is undefined for $x_m=\pm x_{r}$. Thus, to avoid the singularity of the problem, one needs to shift the limit of associated measurement position by $\pm\Delta$ amount ($ \Delta \rightarrow 0 $), whenever required. We encounter such singularity problem only in the case of $-\langle W \rangle$ and $\langle I \rangle$ calculation in entropy dominated limit ($G/D\ll1$). In all other scenarios, this singularity problem does not arise. Therefore, we set the extreme points as the concerning limits to make calculation easier (without any estimation error). Next, we vary the feedback location of the confinement centre ($x_f$) within the range of $0 < x_f < (x_r+x_m)$. In principle one can choose $x_f$ in the other side of the confinement as well, i.e.; $0 > x_f > -(x_r+x_m)$. However, because of the reflection symmetry of the confined structure, the effective potential, and hence, the amount of extractable work is identical for $x_f=\pm x^{\prime}$. Finally, we assumed that the boundary walls are non-interactive and do not exert any force on the particle during a collision. In this regard we mention that  the particle may hit the wall sometime during the feedback protocol, (like: for  $x_f>2x_m$ and when $x_f>x_m+x_r$). We assume that the ‘hitting’ incidents are weak and can not change the temperature of the heat reservoir. However, these hitting incidents can only provide a transient effect on the dynamics of the particle and can not  alter the $P_{eq}(x)$. Therefore, such 'hitting' events do not affect the estimation of $\langle I \rangle $ or $\langle I_u\rangle$. 

\subsection{\label{app:subsec}Numerical simulation details}
We understand that the estimation of most of the physical observable under consideration involves the calculation of equilibrium probability distribution function ($P_{eq}(x)$). We solve the Langevin dynamics (Eqs.~(\ref{meq1}-\ref{meq2})) inside the boundary walls using an improved Euler method \cite{hjorthcomputational} with a time step ($\Delta t = 10^{-3}$) to find a two dimensional probability distribution in a long time ($p(x, y, t\rightarrow \infty )$). We consider a reflecting boundary condition near the confinement walls and employ a Box-Muller algorithm to generate the required thermal noise \cite{Box1958ams}. We obtain $P_{eq}(x)$ by calculation the marginal equilibrium distribution function using Eq.~(\ref{meq7}). For this purpose, we use a spatial grid size of $10^{-1}$ units. We generate a large number of trajectories ($\sim 10^{7}$) to obtain a smooth distribution function. To perform numerical integration, we use a trapezoidal rule with a grid size of $10^{-3}$, whenever it is required. Unless mentioned otherwise, we set $a=0.1$, $c=1.6$ and $D=1$ throughout the manuscript.

\section{RESULTS AND DISCUSSION}
\subsection{ \label{app:subsec} Testament to the Ficks-Jacob's approximation (FJA):}

The amount of extracted work, an average displacement per cycle, and the efficacy are key physical outcomes of the GBIE. As evident from definitions (Eq.~\ref{feq2}-\ref{feq6}), we require to assess the equilibrium probability distribution ($P_{eq}(x)$) of the unshifted confinement ($\lambda(\tau) =0$) for the theoretical estimation of these observable. One can obtain the $P_{eq}(x)$ by numerically solving the  underlying 2D Langevin dynamics (Eq.~(\ref{meq1}-\ref{meq2})) as mentioned earlier. For an analytical estimation of $P_{eq}(x)$, we make use of the equilibrium solution of the Smoluchowski equation (Eq.~\ref{meq8}) in reduced dimension and can written as \cite{Ali2021jcp}:
\begin{equation}\label{19}
    \begin{aligned}
         P_{eq}(x)= N\exp\left[\frac{-A(x)}{D}\right], \\ 
        \text{where} \;\;\;\;  N^{-1} = \int_{-x_r}^{x_r} dxP_{eq}(x)
    \end{aligned}
\end{equation}
where $N$ is the normalization constant. Using Eq.~\ref{meq10} and Eq.~\ref{19}, one can find the $P_{eq}(x)$ under different extent of entropic control:
\begin{equation}\label{20}
\begin{aligned}
 P_{eq}(x) &= \sqrt{\frac{Ga}{\pi D}}\exp\left(-\frac{Ga}{D}x^2\right),  \;\;\;\; \text{for} \;\; \frac{G}{D} \gg 1,\\
 &= \frac{3}{4}\sqrt{\frac{a}{c^3}}(-ax^2+c),  \;\;\;\; \text{for} \;\; \frac{G}{D} \ll 1.
\end{aligned}
\end{equation}
 It must be noted that an assumption of a fast local equilibrium along the transverse direction is necessary for mapping the original two-dimensional Fokker-Planck description (Eq.~\ref{meq5}) into a reduced one-dimensional Smoluchowski equation (Eq.~\ref{meq8}). Therefore, the applicability of the theoretically obtained $P_{eq}(x)$ (Eq.~\ref{19}-\ref{20}) is subjected to the validity of Ficks-Jacob approximation in the considered parameter space.
 \begin{figure}[!ht]
    \centering
    \includegraphics[width=0.45\textwidth]{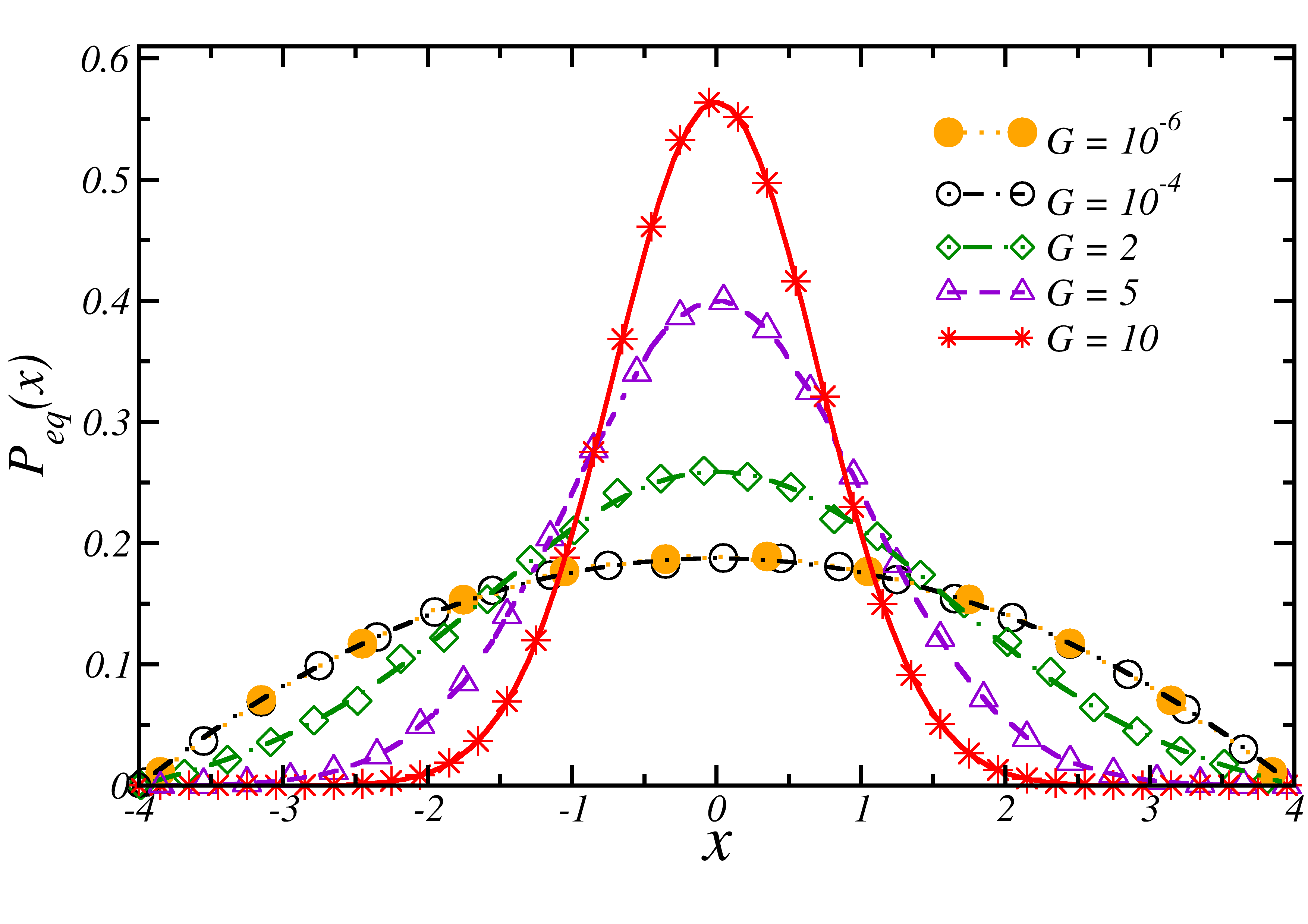}
    \caption{Variation of equilibrium probability distribution function $P_{eq}(x)$ with the position $x$ for different values of transverse bias force $G$. Points are obtained from the numerical simulation study (using Eqs.~\ref{meq1}-\ref{meq2}) and the lines represent theoretical predictions (using Eq.~(\ref{19})). Parameter set chosen: $D=1$, $a=0.1$ and $c=1.6$ for all cases.}
    \label{f3}
\end{figure}
 In Fig.~\ref{f3}, we outline the variation of the initial steady-state probability distribution function ($P_{eq}(x)$) for different transverse force $G$. Numerical integration of Eq.~\ref{meq9} and \ref{19} provides the theoretical predictions for arbitrary value of $G$. Two limiting conditions in transverse force, i.e., $G/D \gg 1$ and $G/D\ll 1$, $P_{eq}(x)$ can be calculated by using Eq.~\ref{20}. All points in Fig.~\ref{f3} correspond to Langevin dynamics simulation data. We obtain a good agreement between the theoretical predictions and numerical simulation data. Thus, it endorses the Ficks-Jacob approximation %as offered to illustrate the dynamics 
 in reduced dimension.\\ 
Fig.~\ref{f3} also depicts that for a high value of external transverse force(here $G = 10$), $P_{eq}(x)$ is a  symmetric Gaussian like function with $\sigma = \sqrt{D/2Ga}$. Where, $\sigma$ is the standard deviation of the probability distribution and can be defined as:
\begin{equation}\label{21}
    \begin{aligned}
     \sigma^2 = \int_{-x_r}^{+x_r}x^2 P_{eq}(x)dx - \bigg(\int_{-x_r}^{+x_r}x P_{eq}(x)dx \bigg)^2.
    \end{aligned}
\end{equation}
In the other extreme ($G \rightarrow 0$), $P_{eq}(x)$ spreads out to a symmetric inverse parabolic function and is independent of $G$. The concerned standard deviation reads as $\sigma=\sqrt{c/5a}$. 

\subsection{Recipe to pull off maximum work \texorpdfstring{$\left ( -\langle W \rangle \right )$}:}

\begin{figure}
    \centering
    \includegraphics[width=0.45\textwidth]{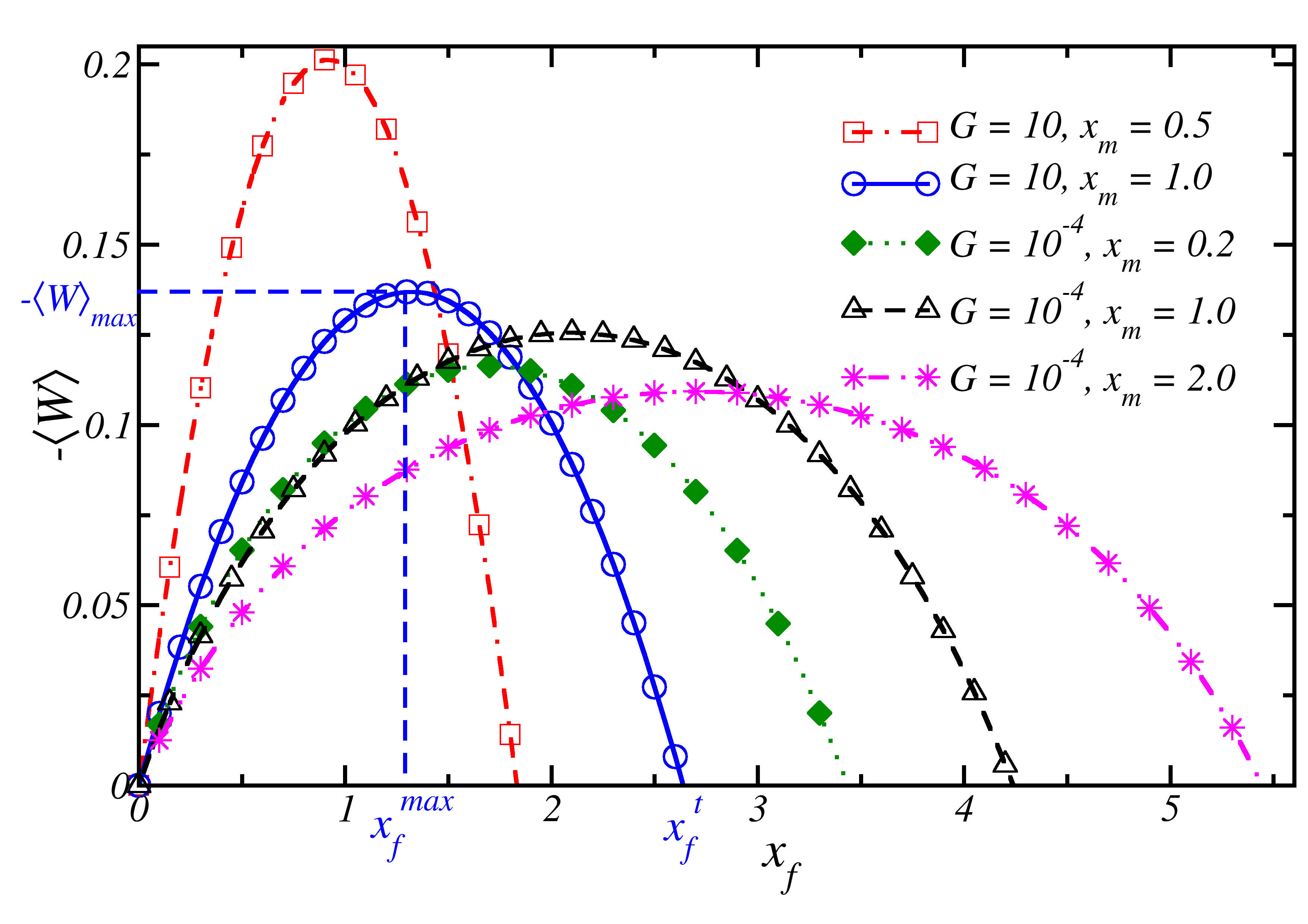}
    \caption{Variation of average work $\left(-\langle W\rangle\right)$ extracted during the feedback cycle with the feedback location  ($x_f$) for different values of measurement position ($x_m$) and different $G$, following Eqs.~\ref{meq10}-\ref{feq2} and \ref{19}. Parameter set chosen: $D=1$, $a=0.1$ and $c=1.6$ for all cases.}
    \label{f4}
\end{figure}

For a given geometric constraint, the measurement distance $x_m$ and the feedback location $x_f$ designate the feedback protocol and hence the outcomes of the GBIE. Also, the dominance of the transverse bias force ($G$)governs the supplemet due to the entropic restraint. Therefore, we examine the adaptation of the averaged work extracted per cycle $-\langle W\rangle$ as a function of the feedback location  $x_f$ for different measurement distances $x_m$ and $G$. Using Eqs.~\ref{meq9}-\ref{feq2} and \ref{19}, one can estimate the average extracted work $-\langle W\rangle$ under any irrational geometric restriction. The outcomes are shown in Fig.~\ref{f4}, and the following observations are perceived:\\\\
(a) For a given measurement distance $x_m$ and transverse force $G$, magnitude of extractable work $\left(-\langle W\rangle\right)$ shows a turnover with the feedback distance $x_f$. A maximum work $\left(-\langle W\rangle_{max}\right)$ can be achieved for an intermediate feedback location, say $x_f=x_{f}^{max}$.\\
(b) The magnitude of $-\langle W\rangle_{max}$ changes with $x_m$ non-monotonically. When all other parameters are kept unchanged, one can realize the highest value of $-\langle W\rangle_{max}$ for an optimum measurement distance. \\ 
(c) One  witness a rise in $-\langle W\rangle_{max}$ with increasing $G$. Thus, the maximum work extraction for a given protocol and confinement parameters is higher in energetic limit than the entropic one.\\
(d) Finally, the monostable geometric trap can serve as an information engine ($ -\langle W\rangle > 0 $) only up to a specific value of the feedback position. We observe that the extraction of work is not possible beyond $x_f > 2x_{f}^{max}$ for any arbitrary system parameter.\\\\
   To apprehend the underlying physics of the observations stated above, we now look into the optimal operating condition on $x_m$ and $x_f$ for maximum work extraction. In the limit of high transverse force ($G/D \gg 1$), one can get the expression of $-\langle W \rangle$ as:
\begin{equation}\label{22}
    \begin{aligned}
     -\langle W\rangle = \sqrt{\frac{GaD}{\pi}}x_f\exp\bigg(-\frac{Ga}{D}x_m^2\bigg)-\frac{Ga}{2}x_f^2erfc\bigg(\sqrt{\frac{Ga}{D}}x_m\bigg),
    \end{aligned}
\end{equation}
where $erfc(z) = 1-erf(z) $ is the complementary error function and  $erf(z) = 2\pi^{-1/2} \int_0^{z} e^{-y^2}dy $.
The solution of $\partial\langle W\rangle/\partial x_f= 0$ with unaltered measurement position $x_m$ yields:
\begin{equation}\label{23}
    \begin{aligned}
     x_f^{max} &= \sqrt{\frac{D}{Ga\pi}}\frac{\exp\bigg(-\frac{Ga}{D}x_m^2\bigg)}{erfc\bigg(\sqrt{\frac{Ga}{D}}x_m\bigg)},
    \end{aligned}
\end{equation}
where, $x_f^{max}$ denotes the feedback location $x_f$ associated to a maximum work extraction.  We have verified the fact that $\partial^2 \langle W\rangle/\partial x_f^2 > 0$ for $x_f=x_f^{max}$ and for any positive values of $G$, $a$ and $D$. Therefore, Eq.~\ref{23} provides the best feedback location to obtain a maximum work in this limit.  The optimal choice of measurement and feedback positions that maximize the $-\langle W \rangle_{max}$ can be obtain by satisfying the $\frac{\partial \langle W\rangle}{\partial x_m} = 0 $ and $\frac{\partial \langle W\rangle}{\partial x_f} = 0$ simultaneously. The exact analytical condition in this limit reads as:
\begin{align}\label{24}
      x_f^* = x_f^{max}|_{x_m=x_m^*}, \;\; \text{and} \;\;\;x_m^* = \frac{x_f^*}{2},
    \end{align}
       where, $x_m^*$ and $x_f^*$ denote the optimal value of measurement and feedback positions, respectively. Plugging them both in Eq.~\ref{23} results in a transcendental equation and can be solved numerically. The solution yields $x_m^*=0.61\sigma$, where $\sigma = \sqrt{D/2Ga}$. Therefore, the observation agrees with the best work extraction restrictions reported earlier \cite{Paneru2018pre,Park2016pre,Paneru2018prl}.\\
     Similarly, under entropic dominance ($G/D \ll 1$), the average work can be calculated as:
    \begin{align}\label{25}
     -\langle W \rangle &= \frac{3}{4}\sqrt{\frac{a}{c^3}}\int_{-x_r+\Delta}^{x_r-\Delta} dx\omega(x)\ln \bigg(\frac{\omega(x-x_f)}{\omega(x)}\bigg)\nonumber\\
    &= T_1(x_m,x_f) + T_2(x_m,x_f) + T_3(x_m,x_f) \nonumber\\ &+ T_4(x_m,x_f)+T_5(x_m,x_f).
     \end{align}
    Where; \begin{align}
   T_1(x_m,x_f) &=  \bigg(\frac{3}{4} \frac{x_f^2}{x_r^2} -\frac{1}{2} \bigg) 
   \ln \bigg| \frac{(x_f+\Delta)(x_r-x_f+x_m)}{(2x_r+x_f-\Delta)(x_r+x_f-x_m)} \bigg|,\nonumber
   \\ T_2(x_m,x_f) &= 
   -\frac{1}{2} \ln \bigg| \frac{2\Delta(x_r-x_f)+x_f x_r}{2x_r(2x_r-\Delta)} \bigg|,
   \nonumber
   \\ T_3(x_m,x_f) &= \frac{a^2x_f}{4c\sqrt{c}}\bigg[(x_r-x_m)(2x_f-x_r+x_m)-2\Delta(x_f-x_r) \bigg],\nonumber
   \\ T_4(x_m,x_f) &= \frac{1}{4}\frac{x_f^2}{x_r^3} \ln \bigg| \frac{a(x_f-x_r)^2 - 2a\Delta(x_f-x_r)-c}{a(x_f-x_m)^2-c} \bigg|,\nonumber
   \\ T_5(x_m,x_f) &= x_m \bigg( \frac{1}{4}\frac{x_m^2}{x_r^2} - \frac{3}{4} \bigg)\ln \bigg| 1 + \frac{ax_f(2x_m-x_f)}{c-ax_m^2} \bigg|\nonumber.
    \end{align}
\begin{figure}
    \centering
    \includegraphics[width=0.45\textwidth]{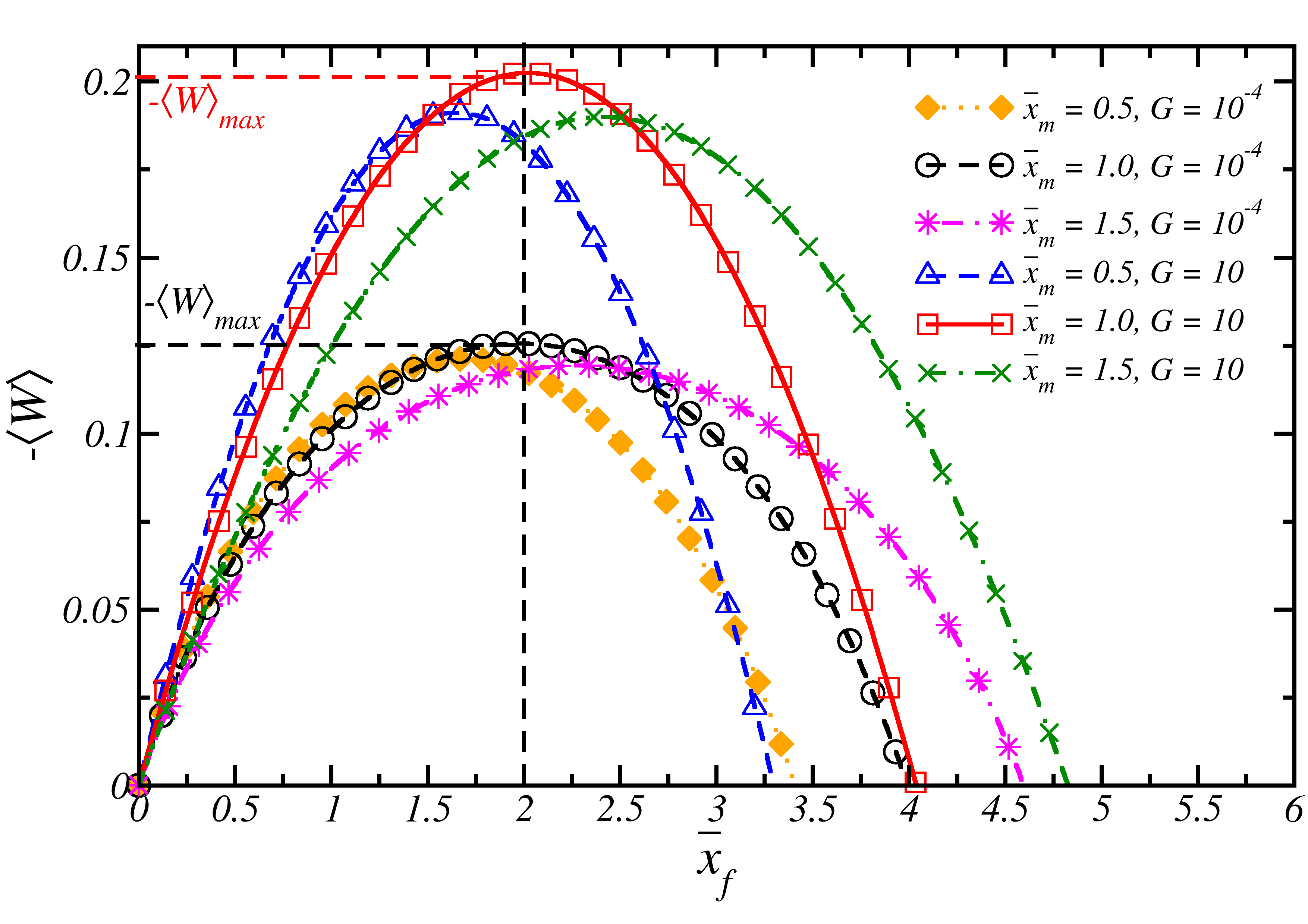}
    \caption{Variation of average work ($-\langle W\rangle$) extracted  with the scaled feedback position ($\bar{x}_f$) for different values of scaled measurement length ($\bar{x}_m$). Orange solid diamond, black circle, and magenta starred points are associated to $\bar{x}_m = 0.5,\; 1.0 $ and $1.5$, respectively in a low $G$ limit ($G=10^{-4}$). Blue triangle, red square and green cross points are associated to $\bar{x}_m =0.5,\; 1.0$ and $1.5$, respectively in an energy dominated condition ($G=10$). Parameter set chosen: $D=1$, $a=0.1$ and $c=1.6$ for all cases.}
    \label{f5}
\end{figure}
Here, $|...|$ denotes the absolute value of the observable. One can derive $x_f^{max}$ theoretically following a similar method  explained for the energy-dominated case. For a given $x_m$, the corresponding $x_f^{max}$ will be  the solution of the following transcendental equation (with $x_f=x_f^{max}$):
\begin{eqnarray}\label{26}
     \Theta_1(x_f)&+&\Theta_2(x_f)+\Theta_3(x_f)=0 \nonumber \\
     \Theta_1(x_f) &=& -x_fx_r \ln \bigg| \frac{ax_f(a(x_m-x_f)+\sqrt{ac})}{(2\sqrt{ac}-ax_f)(a(x_m-x_f)-\sqrt{ac})} \bigg|\nonumber\\
     \Theta_2(x_f) &=& \frac{1}{2} (x_m-x_r) (x_m+x_r+2x_f ) \nonumber\\ 
     \Theta_3(x_f) &=& \frac{-x_f^2}{2} \ln \bigg| \frac{2x_f\sqrt{ac}-ax_f^2}{ax_m^2-2ax_fx_m+ax_f^2-c} \bigg|.  
 \end{eqnarray}
  From Eqs.~\ref{25}-\ref{26}, one obtains the best work extraction requirement as: $x_f^\ast = x_f^{max}|_{x_m=x_m^\ast}$ and $x_m^\ast = \frac{x_f^\ast}{2}$. The condition yields:

    \begin{align}\label{27}
 2x_m^\ast \ln  \bigg|\frac{-ax_m^{\ast ^2}+c}{4x_m^\ast (-ax_m^\ast +\sqrt{ac})} \bigg| + 2x_m^\ast x_r \ln \bigg |1+\frac{x_r^2}{x_m^\ast } \bigg | \nonumber \\ + \frac{1}{2}(x_m^\ast -x_r)(5x_m^\ast +x_r) = 0.
    \end{align}
 Solution of the transcendental equation Eq.~\ref{27} gives the best recipe as, $x_m^*=0.6\sigma$ and $x_f^*=2x_m^*$, where the standard deviation in this limit reads as $\sigma = \sqrt{c/5a}$.\\
 
 We find that for $G=10$, the maximum work is obtained when $x_m\approx 0.42$ and $x_f \approx 0.84$, and $\sigma \approx 0.71$. In the limit of $G \rightarrow 0$, $x_m\approx 1.07$ and $x_f \approx 2.14$, with $\sigma \approx 1.79$ provides the optimal condition for maximum work extraction. Therefore, despite of the differences in dominance of $G$, the recipe to obtain maximum work remains same as $x_m = 0.6\sigma$ and $x_f=2x_m$. We revisit the variation of the average work extracted per cycle $-\langle W\rangle$ as a function of a scaled position of the shifted confinement $\bar{x}_f$ for different scaled measurement distance $\bar{x}_m$ and $G$. Here, we define a scaled observable $\bar{R}$ as $\bar{R}=R/x_m^*$. Results are shown in Fig.~\ref{f5}.\\
 
 Fig.~\ref{f5} clearly shows that $-\langle W\rangle_{max} $ is maximum for $\bar{x}_f=2$, irrespective to the values of $G$. We determine the $x_f^{max}$ for a given $x_m$ using Eq.~\ref{23} and \ref{26} in respective limits of $G$. In this context, we mention that one can verify the aforementioned relation for any arbitrary values of $G$ by estimating direct numerical integration of Eq.~\ref{feq1}-\ref{feq2}, \ref{19} and \ref{21}. As mentioned earlier, Fig.~\ref{f4}-\ref{f5} also shows that the geometric trap can extract work ($ -\langle W\rangle > 0 $) only up to a certain value of the feedback position. The restriction $ -\langle W\rangle = 0 $ in Eq.~\ref{22} (under constant $x_m$) gives the upper bound of the feedback location as $x_f^{t}$. Using Eq.~\ref{22} and \ref{25} , one can show that $x_f^{t}=2x_f^{max}$ invariant to the extent of entropic dominance.\\
 
To shine more light on these observations and to examine the differences of $-\langle W \rangle_{max}$ between high and low values of $G$, we investigate the amount of total information ($\langle I \rangle$) and unavailable information ($\langle I_u \rangle$) during the feedback protocol.
\begin{figure}
    \centering
    \includegraphics[width=0.45\textwidth]{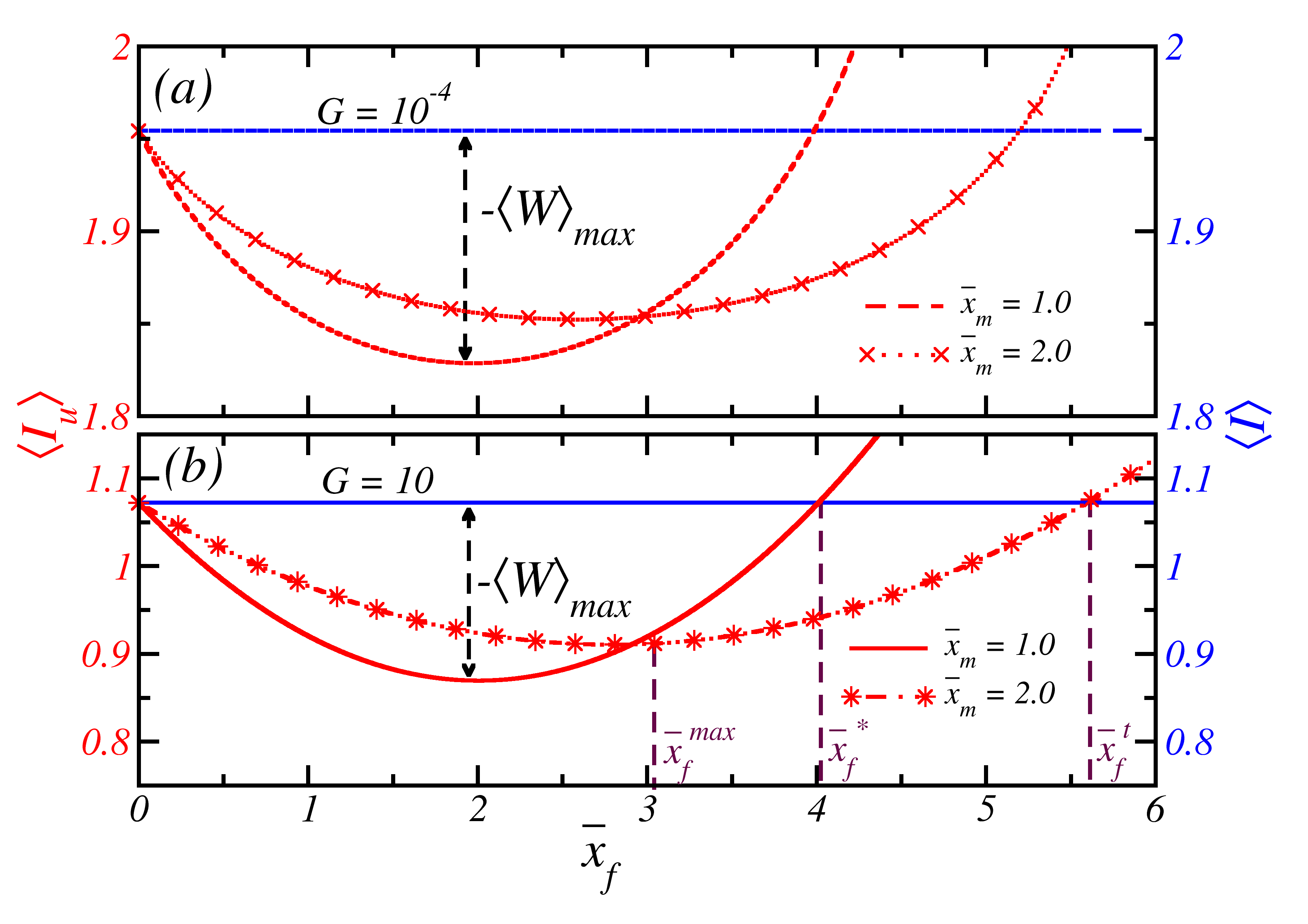}
    \caption{Change in average Information ($\langle I\rangle$) and unavailable  Information ($\langle I_u\rangle$) with increasing scaled feedback position $\bar{x}_f$ for different values of scaled measurement position ($\bar{x}_m$). Parameter set chosen: $D=1$, $a=0.1$ and $c=1.6$ for all cases. (a) Represents the variation in entropy dominated regime $G=10^{-4}$. (b) The same in energy controlled situation $G=10$.}
    \label{f6}
\end{figure}
In the limit of $G/D \gg 1$, the net information acquired by the measurement can be estimated using Eq.~\ref{feq3} and Eq.~\ref{20}:
\begin{equation}\label{28}
    \begin{aligned}
    \langle I \rangle \simeq \frac{1}{2} - \ln \bigg( \sqrt{\frac{Ga}{\pi D}} \bigg)
    \end{aligned}
\end{equation}
Similarly, for the other extreme $G/D \ll 1$, the net information has the form (using Eq.~\ref{feq4} and Eq.~\ref{20}):
\begin{equation}\label{29a}
    \begin{aligned}
    \langle I \rangle \simeq \frac{5}{3} - \ln \bigg( 3\sqrt{\frac{a}{c}}\bigg), \;\;\text{in the limit of $\Delta \rightarrow 0$} . 
    \end{aligned}
\end{equation}
For any arbitrary values of $G$, $x_m$ and $x_f$, using Eqs.~(\ref{feq1}-\ref{feq4}) and Eqs.~(\ref{19}-\ref{20}) one can show that:
\begin{equation}\label{30}
 D (\left \langle I \right \rangle- \left \langle I_u \right \rangle)= -\left \langle W \right \rangle.
\end{equation}

Before we advance further, we mention that the Eq.~\ref{30} signifies that for an instantaneous (error-free) measurement and feedback process, available information acquired during the protocol has entirely been extracted as work \cite{Paneru2018pre,Park2016pre,Paneru2018prl}. Popularly, such types of engine are denoted as lossless information engines \cite{Paneru2018prl}. 
Therefore, the GBIE can be considered as lossless engine in the sense that it converts entire  available information $(\left \langle I \right \rangle- \left \langle I_u \right \rangle))$ into work. In a true sense, however, it is not completely lossless. Because of the irreversible nature of the protocol, a finite amount of acquired information is lost during the relaxation phase. However, it is worthwhile to mention that one may design and introduce a reversible protocol in which the net acquired information is equivalent to available information ($\left \langle I_u \right \rangle=0$) \cite{dinis2020entropy,granger2016}.

In Fig.~\ref{f6}, we study the variation of average Information ($\langle I\rangle$) and unavailable  Information ($\langle I_u\rangle$) during the feedback with the scaled position of the shifted confinement centre $\bar{x}_f$ for different values of $\bar{x}_m$. Results show that the total acquired information is independent of feedback position for a given system parameter set. However, the amount of unavailable information $(\left \langle I_u \right \rangle)$ varies non-monotonically with increasing $x_f$. In the limit of $x_f \rightarrow 0$, all the acquired information is lost during the relaxation process as there is no change in the effective potential of the system. Using Eq.~\ref{feq3}-\ref{feq4}, one realize $\langle I_u\rangle \simeq \langle I\rangle$ for $x_f \rightarrow 0$.
With an increasing $x_f$, the distance between the measurement position and the feedback site decreases. Consequently, the number of singular paths decreases during relaxation and particles reach the new potential minimum with less uncertainty. This results in a decrease in unavailable information. As a result, the amount of extractable work increases.
At this stage it is worthwhile to mention that the loss of acquired information during measurement happens because of the presence of {\it unusual} pathways (singular) during the relaxation process. For the protocol with error-free measurement, a measured outcome is greater (or lesser) than $x_m$ if and only if the particle resides in the $x>x_m$ (or $x<x_m$) region.
We then employ the feedback based on the measurement outcome and allow the system to relax. However, after the relaxation stage, particles can reside in the $x>x_m$ (or $x<x_m$) region even though the measured outcome was greater (or lesser) than $x_m$. We denote these relaxation pathways as singular paths; the measurement of such paths contributes to information lost during the process. 
For a better understanding of singular paths and their contribution in determining $\langle I_u \rangle$, we refer to Szilard's engine as discussed in \cite{Ashida2014pre}.

Coming back to Fig.~\ref{f6}, for $x_f \gg x_m$, $\langle I_u\rangle > \langle I\rangle$ as the contribution of the second term of the right-hand side of Eq.~\ref{feq4} dominates. The situation corresponds to a large shift in the detention centre (hence the effective potential). The distance between $x_m$ and $x_f$ becomes large again. Thus, the number of singular paths during the relaxation processes increases again, resulting in a decrease in the magnitude of $\langle W \rangle$ due to the heavy information loss during relaxation. Therefore, one optimum $x_f$ distance exists where the loss of information is minimum, and the condition corresponds to the recipe of maximum work extraction. For a given protocol the $\langle I\rangle$ is invariant to the feedback position (see Eqs.~\ref{feq3} and \ref{28}-\ref{29a}). Therefore, the relation \ref{30} suggests that the condition for a minimum $\langle I_u\rangle$ is identical to the optimal recipe of maximum work extraction.

The argument promotes the existence of a nonzero $x_f^t$  for which total information levels the loss due to the relaxation process. Beyond this point, the protocol results in refrigeration (-$\langle W \rangle < 0$). Finally, Fig.~\ref{f6} also shows that both the information and unavailable information increase for decreasing $G$. However, in the low $G$ limit, the rise in unavailable information is relatively higher than the total information. Consequently, one can witness that the amount of maximum extractable $-\langle W \rangle_{max}$ at optimal measurement distance $x_m^*$ is higher in energy ruled region than that of the entropic case.
\begin{figure}
    \centering
    \includegraphics[width=0.45\textwidth]{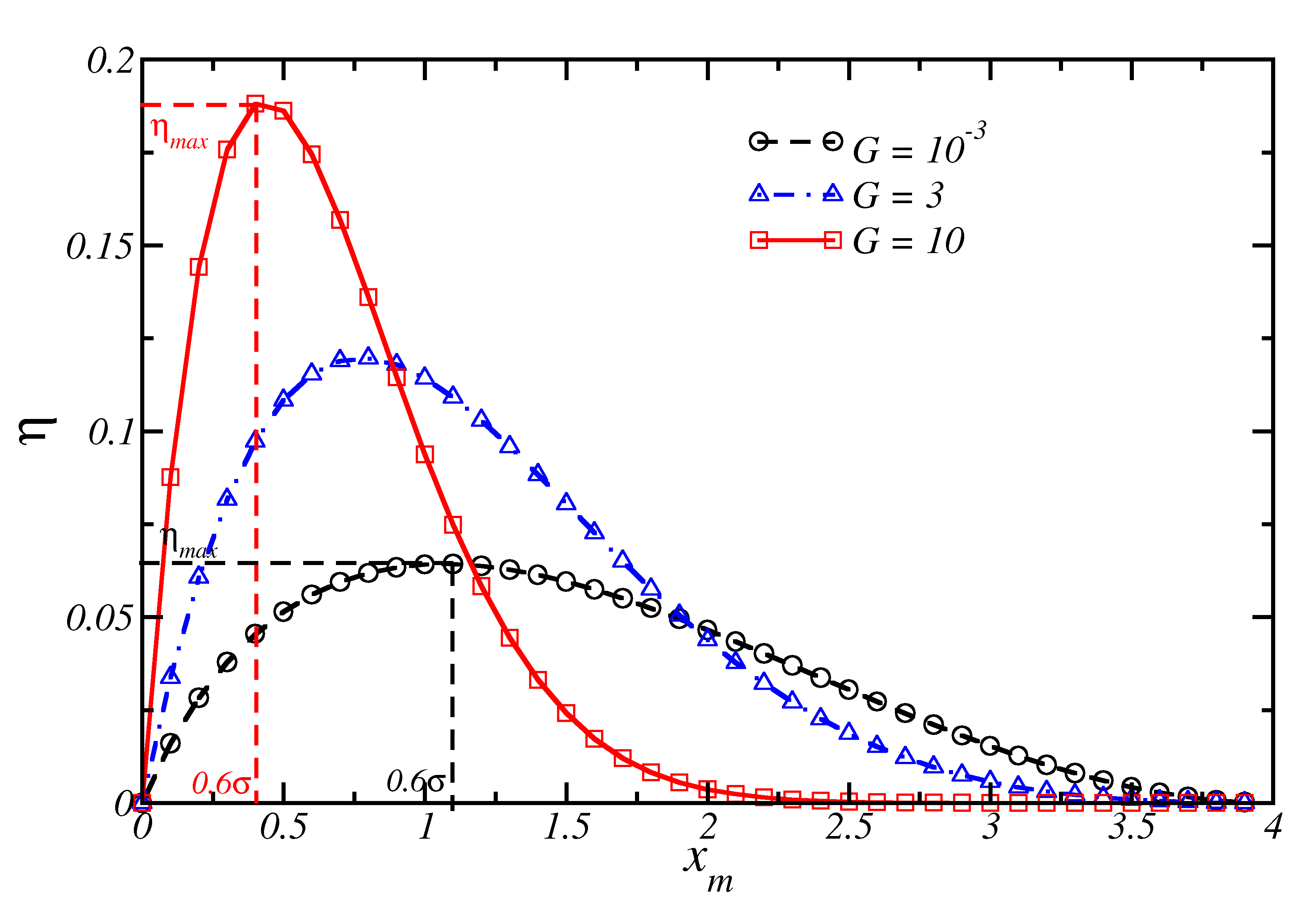}
    \caption{Variation of the efficiency ($\eta$) with the measurement position ($x_m$) for a fixed feedback location  ($x_f = 2x_m$) and different $G$. Parameter set chosen: $D=1$, $a=0.1$ and $c=1.6$ for all cases.}
    \label{f7}
\end{figure}

 Next, we calculate the efficiency of the information engine, which can be defined as $\eta = -\langle W \rangle / D\langle I \rangle $. In Fig.~\ref{f7}, we study the variation of efficiency ($\eta$) with the measurement distance ($x_m$), for the protocol with a feedback site twice of measurement distance ($x_f = 2x_m$). Results show that the engine efficiency varies non-monotonically with the measurement distance ($x_m$). The variation depicts that the engine's efficiency is always less than unity, and it has a maximum at the measurement distance  $ \approx 0.6 \sigma$ ($\sigma$ is the standard deviation) irrespective of the magnitude of the transverse force. As the efficiency can not reach unity, the engine is not a completely lossless one. Also, the engine is most efficient when unavailable information is minimal during the employed feedback process. Fig.~\ref{f7}, it is evident that the $\eta_{max}$ decreases with higher entropic control of the system. This reduction in $\eta_{max}$ can be attributed to the lower work extraction and rise in the available information in the low $G$ limit.

Finally, one can find that the present set-up is consonant with the integral fluctuation theorem \cite{Ashida2014pre}:
\begin{equation}\label{31}
\begin{split}
   &\left \langle \exp\left(-\frac{W_d}{D}-I+I_u \right) \right\rangle=\int_{-x_r}^{x_m} {dx} P_{eq}({x})\\
   &+\int_{x_m}^{x_r} {dx} P_{eq}({x})\exp\left(\frac{A(x)-A(x-x_f)}{D}\right)\frac{P_{eq}({x})}{P_{eq}(x-x_f)}= 1.
    \end{split}
\end{equation}
As the associated free energy change is zero ($\Delta F=0$), we estimate the dissipated work as $W_d=A(x-x_f)-A(x)$.

\subsection{Optimization of average displacement ($\Delta x$) and efficacy ($\gamma$):}
\begin{figure}
    \centering
    \includegraphics[width=0.45\textwidth]{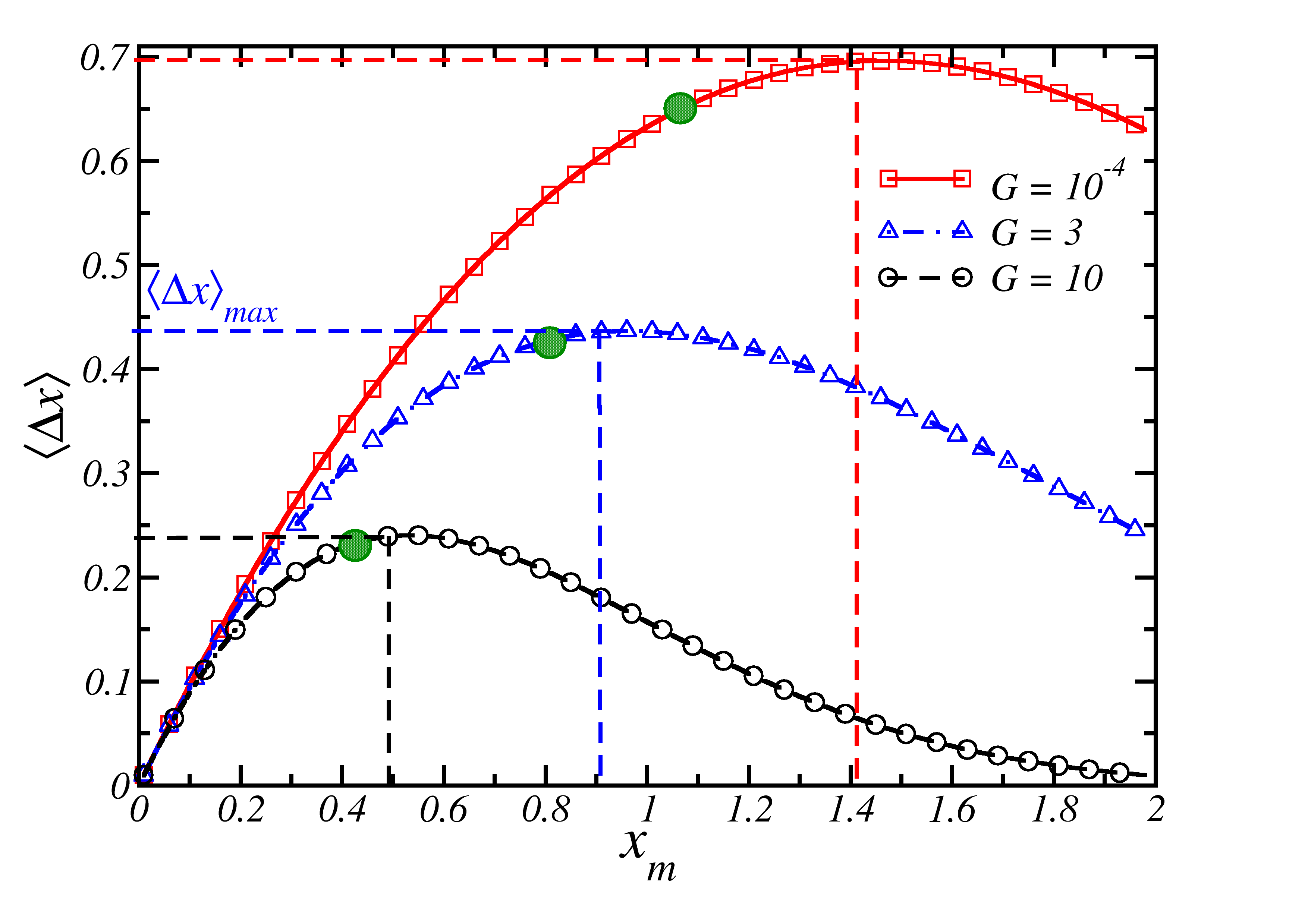}
    \caption{Variation of the average unidirectional step per cycle $\left(\langle \Delta x\rangle\right)$ observed during the feedback protocol with the measurement position ($x_m$) for different values of downwards bias force ($G$). Parameter set chosen: $x_f=2x_m$, $D=1$, $a=0.1$ and $c=1.6$ for all cases. Green colored filled circles indicate the $\langle \Delta x\rangle$ associated to the best work extraction condition ($x_m^*\sim 0.6\sigma$). }
    \label{f8}
\end{figure}

The average displacement per cycle ($\langle \Delta x \rangle$) measures the mean unidirectional motion of particles during the feedback mechanism. Therefore, $\langle \Delta x \rangle$ quantifies unidirectional transport induced by the information engine that operates in a single heat bath. For an irrational choice of the transverse bias force ($G$), we obtain $\langle \Delta x \rangle$ using Eq.~\ref{feq6}, and Eq.~\ref{19}. The definition of $\langle \Delta x \rangle$ (Eq.~\ref{feq6}) clearly shows that it varies linearly with respect to the feedback position ($x_f$). Inspired by the observation of the previous subsection, one can assume that a good feedback location depends on the choice of the measurement distance.  Therefore, we consider a fixed feedback distance, as double of the measurement position $x_f=2x_m$, and examine the response of $\langle \Delta x \rangle$ for different $x_m$. Fig.~\ref{f8} shows the variation of $\langle \Delta x \rangle$ as a function of the  measurement position ($x_m$) for different entropic control. The variations depict a turnover in $\langle \Delta x \rangle$ with increasing $x_m$. Eq.~\ref{feq6} under the constraint of  $x_f=2x_m$ indicates a trade-off between the $x_m$ and integrated marginal probability ($P(x,t)$) of particles beyond $x_m$ in determining $\langle \Delta x \rangle$.  To obtain a quantitative measure of the optimal control on $\langle \Delta x \rangle$, we recall Eq.~\ref{feq6} and Eq.~\ref{20} and examine the limiting responses. Under the restriction of  $x_f=2x_m$, we find:
\begin{equation}
    \begin{aligned}\label{31a}
     \langle \Delta x \rangle &= x_m erfc \bigg ( \sqrt{\frac{Ga}{D}} x_m \bigg ) \;\;\;\; for\; \frac{G}{D} \gg 1, \\ &= x_m - \frac{3}{2}\sqrt{\frac{a}{c}}x_m^2 + \frac{1}{2}\sqrt{\frac{a^3}{c^3}}x_m^4  \;\;\;\; for\; \frac{G}{D} \ll 1 .
     \end{aligned}
\end{equation}
Thus, in the both ends of $G$, $\langle \Delta x \rangle$ varies non-monotonically with $x_m$. This drives the manifestation of an optimum measurement distance to accomplish the largest average displacement per cycle.\\

Fig.~\ref{f8} also reveals that both  the best average distance $\langle \Delta x \rangle_{max}$ and concerned measurement distance increases by introducing more entropic control to the process. 
One can determine the optimum value of $x_m$ for achieving maximum $\langle \Delta x \rangle$ by maximising Eq.~\ref{31a}.
For a high $G$, $\frac{\partial \langle \Delta x \rangle }{\partial x_m} = 0$  yields:
\begin{equation}\label{29}
    \begin{aligned}
     x_m = \sqrt{\frac{\pi}{2}}\sigma erfc\bigg(\frac{x_m}{\sqrt{2}\sigma} \bigg){\exp\bigg( \frac{-x_m^2}{2\sigma^2} \bigg)}.
    \end{aligned}
\end{equation}
Numerical solution of this transcendental equation gives $x_m=0.75\sigma$, where $\sigma=\sqrt{D/2Ga}$.
For $G \rightarrow 0$, the restriction $\frac{\partial \langle \Delta x \rangle}{\partial x_m} = 0 $ results in:
\begin{equation}\label{35}
    \begin{aligned}
     1-\frac{3}{\sqrt{5}}\frac{x_m}{\sigma} + \frac{2}{5\sqrt{5}}{\frac{x_m^3}{\sigma^3}} = 0.
    \end{aligned}
\end{equation}
The solution of the cubic polynomial gives the measurement position related to best average displacement as $0.81\sigma$, where the standard deviation $\sigma=\sqrt{c/5a}$ for this case. Finally, it is noteworthy that the requirement to have a maximum  $\langle \Delta x \rangle$ is not identical to the best work extraction condition. Using these control we find the best average displacement $\langle \Delta x \rangle \approx 0.24$ and $\approx 0.70$ for $G=10$ and $G=10^{-4}$, respectively. Therefore, the maximum average displacement per cycle is higher under an entropic control than in the energy governed set-up. %The observation can be visualized in Fig.~\ref{f7}. 
One can explain the enhanced value of $\langle \Delta x \rangle_{max}$ for a pure entropic information engine in terms of the shape of the equilibrium probability distribution. In the limit of $G\rightarrow 0$, the $P_{eq}(x)$ has an inverted parabola like outline along the feedback coordinate (Fig.~\ref{f3}) . This increases the standard deviation of the distribution in comparison to the $P_{eq}(x)$ with high transverse bias. Consequently, the marginal probability towards the confinement terminus ($x_m < x < x_r$) is higher under entropic control. Therefore, the integrated probability of particles crossing a high measurement distance is higher. This results in high $\langle \Delta x \rangle$ in a pure entropic GBIE.\\
\begin{figure}
    \centering
    \includegraphics[width=0.45\textwidth]{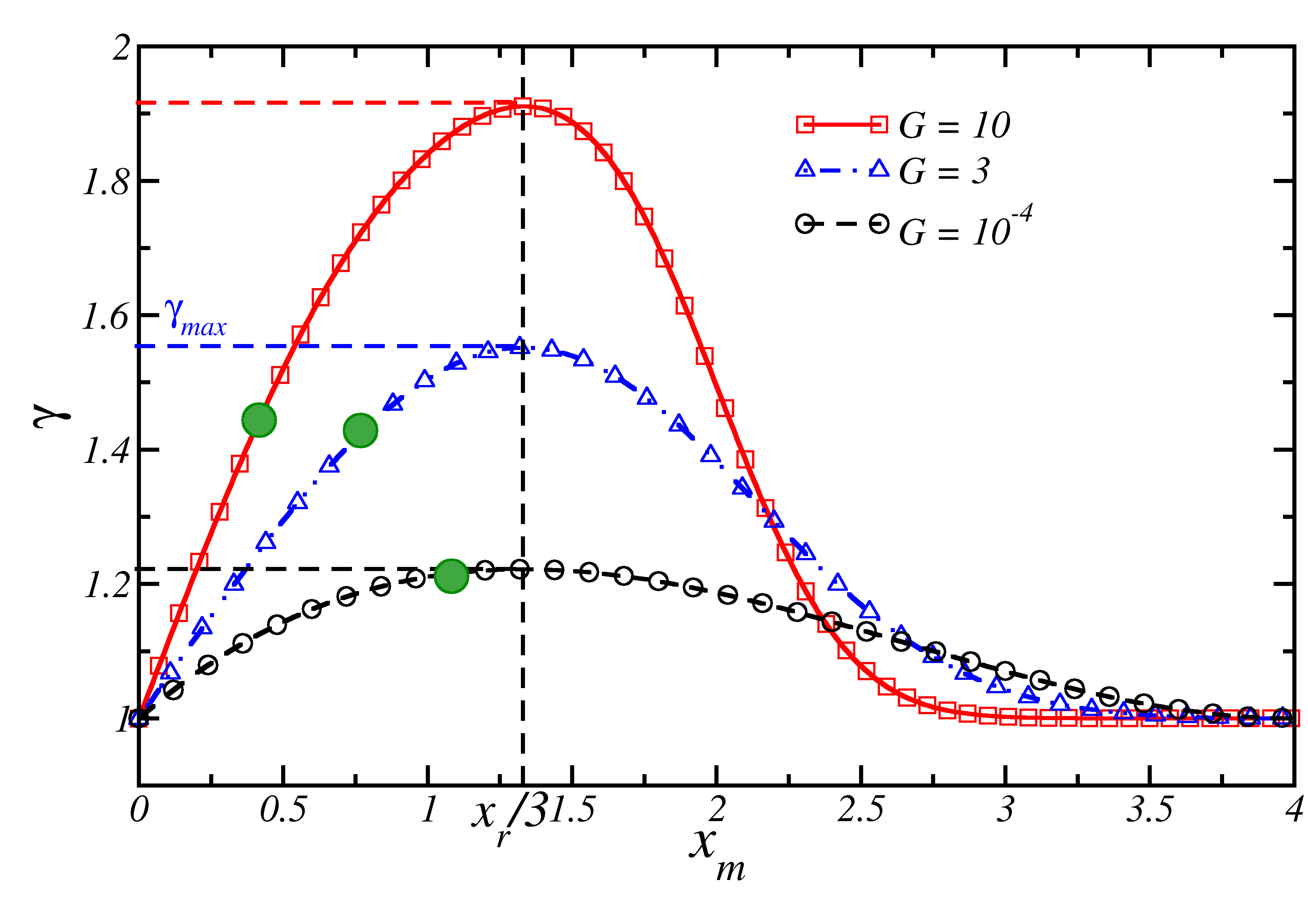}
    \caption{Variation of the efficacy of the feedback protocol ($\gamma$) with the measurement position ($x_m$) for different $G$. Parameter set chosen: $x_f=2x_m$, $D=1$, $a=0.1$ and $c=1.6$ for all cases. Green colored filled circles indicate the efficacy associated to the best work extraction condition ($x_m^*\sim 0.6\sigma$).}
    \label{f9}
\end{figure}

Finally, we study the hallmarks of the efficacy of the feedback controller of a GBIE, using Eq.~\ref{feq5} under the condition of $x_f=2x_m$. The results are shown in Fig.~\ref{f9}. Following observations are evident. The efficacy ($\gamma$)  varies non-monotonically with increasing measurement position. 
The efficacy approaches to unity, $\gamma = 1$ for both the extreme of the measurement distances, i.e.; either $x_m = 0$ or $x_m \rightarrow x_r$ for all three different values of transverse force $G$. The magnitude of the maximum efficacy $\gamma \approx 1.9$ is higher for $G=10$, whereas it  is less for $G = 10^{-4}$, $\gamma \approx 1.22$. The corresponding measurement position to maximum efficacy is obtained at $x_m \approx 1.33$, invariant in $G$. Thus, the recipe to have $\gamma_{max}$  differs to the best work extraction prescription. 

Using Eq.~\ref{feq5} and ~\ref{20}, the efficacy of the process under energetic and entropic extreme takes the form:
\begin{equation}\label{31}
   \begin{aligned}
    \gamma = \frac{1}{2} + \frac{1}{2} \bigg[erf \bigg( \sqrt{\frac{Ga}{D}}x_m \bigg) &- erf \bigg( \sqrt{\frac{Ga}{D}}(x_m-x_f) \bigg) \\ &+ erf \bigg( \sqrt{\frac{Ga}{D}}(x_r-x_f) \bigg) \bigg],
   \end{aligned}
\end{equation}
and
\begin{equation}\label{32}
    \begin{aligned}
     \gamma = 1 + \frac{3}{4} \sqrt{\frac{a^3}{c^3}} x_f (x_r - x_m) (x_r + x_m - x_f ),
    \end{aligned}
\end{equation}
respectively. In either cases, the efficacy of the engine, under the constrain $x_f = 2x_m$, reduces to a uni-variate function of $x_m$. In the limit of $G/D >> 1$, Eq.~\ref{31} reduces to
\begin{equation}\label{33}
    \begin{aligned}
    \gamma = \frac{1}{2} +  erf \bigg( \sqrt{\frac{Ga}{D}}x_m \bigg) + \frac{1}{2} erf \bigg( \sqrt{\frac{Ga}{D}}(x_r-2x_m) \bigg ).
    \end{aligned}
\end{equation}
On the other hand, the restriction $x_f = 2x_m$ reduces Eq.~\ref{32} to:
\begin{equation}\label{34}
\begin{aligned}
\gamma  = 1 + \frac{3}{4} \sqrt{\frac{a^3}{c^3}} 2x_m (x_r - x_m)^2.
\end{aligned}
\end{equation}
  
\begin{figure}
    \centering
    \includegraphics[width=0.5\textwidth]{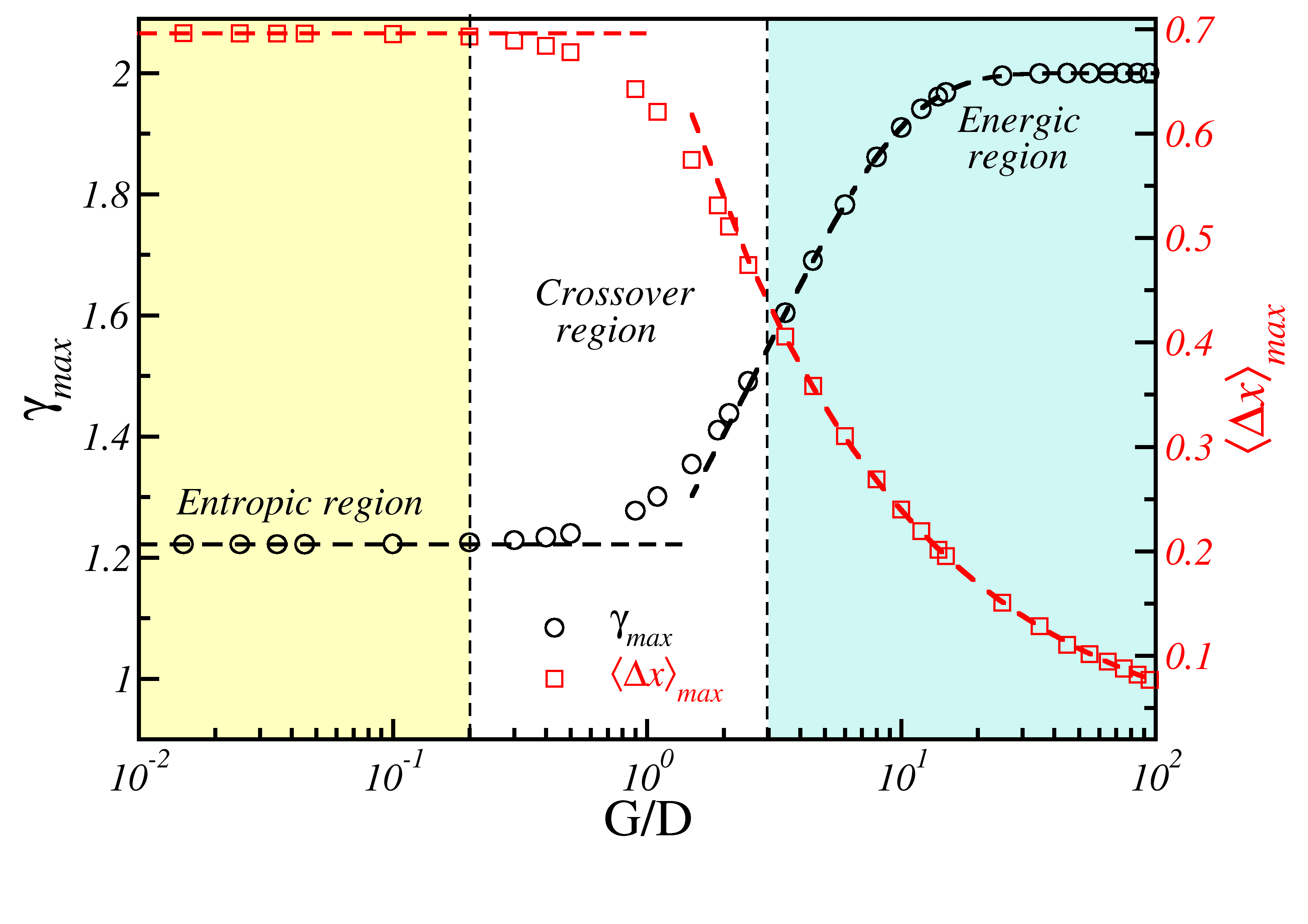}
    \caption{Variation of the best efficacy ($\gamma_{max}$) and the best displacement per cycle $\langle \Delta x \rangle_{max}$ with  transverse force ($G/D$). Parameter set chosen: $x_f=2x_m$, $D=1$, $a=0.1$ and $c=1.6$ for all cases. Points are obtained by the numerical integration of the general expressions of efficacy (Eq.~\ref{feq5}) and averaged distance per cycle (Eq.~\ref{feq6}) under the restriction of $x_f=2x_m$. Dashed lines are obtained from the theoretical expressions related to the limiting behavior.}
    \label{f10}
\end{figure}

From, Eq.~\ref{33} and \ref{34}, it is obvious that  $\gamma$ converges to unity for either extreme of the measurement positions, $x_m = 0$ and $x_m \rightarrow x_r$, irrespective to the strength of the transverse force.  
Also, as expressed in Eq.~\ref{33}-\ref{34}, $\gamma$ varies non-monotonically with $x_m$. %One can find the optimum value of $x_m$ that corresponds to a maximum efficacy imposing the restriction $\frac{\partial \gamma}{\partial x_m} = 0$ on Eq.~\ref{32}. 
We find that the best value of $x_m$ that generates maximum efficacy is $\frac{x_r}{3}$ in either case and is independent of $G$ and any other geometric parameter. Finally, using  the condition $x_m = \frac{x_r}{3}$ on  Eqs.~\ref{33} and \ref{34}, one can  find that $\gamma \rightarrow 2$ in the limit of  $G/D >> 1$, whereas $\gamma =\frac{11}{9}$ in the other extreme. As mentioned earlier, the spread of the marginal probability distribution is broader in an entropy dominated situation, and hence, the particles can relax in a higher number of paths. Therefore, the protocol's efficacy reduces compared to an energetic system. To summarize the subsection, we depict that both $\langle \Delta x \rangle_{max}$ and  $\gamma_{max}$, show a cross-over response once the system is driven from an entropic to an energetic dominated regime, as shown in Fig.~\ref{f10}.

Before we conclude, we mention a few pertinent observations on the experimental feasibility of a GBIE. The set-up requires a Brownian diffusion inside a narrow channel with irregular geometry and suitable measurement techniques that can relate the underlying available information to the thermodynamic outputs. A recent experiment on diffusion through a corrugated channel by Yang et al. demonstrates an entropy-driven transport \cite{yang2017}. They have fabricated irregular channels using a two-photon writing system followed by the imaging procedure and studied the diffusion of fluorescently labelled polystyrene colloidal particles inside the cavity. Furthermore, the study also validates the Ficks-Jacobs approximation once the hydrodynamics effects are considered. In another investigation, researchers studied the entropic ratcheting effect due to channel asymmetry \cite{marquet2002}. One can microfabricate a narrow channel using photolithography \cite{marquet2002}. The diffusion of colloids across a constrained geometry has been studied using microfluidics and holographic optical tweezers \cite{pagliara2014,pagliara2014prl}. On the other hand, recent experimental developments illustrate the design principle of different Brownian information engines \cite{Berut2012nat,Paneru2018prl,Paneru2018pre,Paneru2020natcommun,Dago2021,Paneru2020}. These studies display the interconversion between the information and other thermodynamic outcomes. Therefore, one can map the entropic constraints by designing a suitable narrow cavity in the spirit of \cite{yang2017,marquet2002,pagliara2014,pagliara2014prl} and can measure the thermodynamic observable ( appeared from available information) introducing feedback procedures employed in \cite{Berut2012nat,Paneru2018prl,Paneru2018pre,Paneru2020natcommun,Dago2021,Paneru2020}. The outcome of the current study, thus,  orchestrates a perfect standard to devise such geometric information engines.

\section{Conclusions}
We explore the optimum operating condition of a GBIE built of Brownian particles trapped in monostable confinement and subjected to error-free feedback regulation. The cycles utilize the information gathered for work extraction and submit a unidirectional passage of the particle. The upshots of the measurement position $x_m$ and the feedback site $x_f$ circumscribe the engine's performance. We determine the optimal condition for maximizing the extracted work ($-\langle W \rangle$),  the average displacement per cycle ($\langle \Delta x \rangle$)  and the effectiveness of the protocol ($\gamma$) under varying entropic authority.
\\

 Analogous to other Brownian information engines \cite{Paneru2018pre,Park2016pre,Paneru2018prl},  the GBIE under feedback controller can completely utilize the available information and hence, be regarded as a lossless information engine. We specify the criteria for utilizing the available information in an output work and the optimum operating requisites for best work extraction. The maximum work extraction is possible when $x_m = 0.6\sigma$ and $x_f = 2x_m$. The observation is consonant with the best work extraction restrictions reported earlier \cite{Paneru2018pre,Park2016pre,Paneru2018prl} and showed the universality of requisites. Nonetheless, the measurement distance and feedback site positions alter upon remodelling of entropic dominance as the standard deviation itself develops during such parameter tuning. In an energy dominated process, $\sigma$ depends on the ratio of the thermal energy to the advective energy of the process as $\sigma = \sqrt{D/2Ga}$. On the other hand,  $\sigma$ becomes independent of $G$ and depends only on the geometric aspect ratio ($x_r=\sqrt{c/a}$) in a purely entropic control.  The magnitude of the extracted work grows with increasing transverse force $G$. One can justify the lower benefits of achievable work in an entropy ruled scenario in terms of the elevated loss in information during the relaxation process.\\
 Next, we find the condition on $x_m$ for maximum average displacement per cycle ($\langle \Delta x \rangle$) with a restriction on the feedback site as $x_f = 2x_m$. The measurement position that gives the best average displacement varies with the extent of entropic control. For high $G$ values ($\gg1$), we find $x_m \sim 0.75\sigma$ and in the limit of $G \rightarrow 0$, the $x_m \sim 0.81\sigma$ is responsible for best unidirectional motion ($\langle \Delta x \rangle_{max}$). Therefore, unlike the work extraction, the mean unidirectional displacement of particles is higher in the entropy-dominated regime than in the energy governed system.
Upon decreasing $G$, the spread of the equilibrium marginal probability distribution ($P_{eq}(x)$) becomes wider (higher $\sigma$). Consequently, a more fraction of particle can satisfy the measurement requirement, which increases the $\langle \Delta x \rangle_{max}$ value in an entropicaly driven system.\\

Finally, under a given feedback location $x_f=2x_m$, the maximum efficacy ($\gamma_{max}$) is invariant with the transverse force strength $G$ and achieved when $x_m = x_r/3$ . $\gamma_{max}$  approaches the universal upper bound $2$ under firm energetic control. On lowering down the energetic power, the upper bound becomes tighter and shows crossover to $\gamma_{max} = \frac{11}{9}$ for a pure entropic device. We trust that the outcomes of the present study will help to design geometric information engines
and will result in new scopes for further theoretical and experimental investigations.

\begin{acknowledgments}
 RR and SYA acknowledge  IIT Tirupati for fellowship. DM thanks SERB (Project No. ECR/2018/002830/CS), Department of Science and Technology, Government of India, for financial support and IIT Tirupati for the new faculty seed grant. 
\end{acknowledgments}
\section*{Data Availability}
The data that support the findings of this study are available within the article.
%\end{Data Availability}
\section*{Conflict of interest}
The authors have no conflicts to disclose.

%\section{Appendixes}

%\subsubsection{\label{app:subsubsec}A subsubsection in an appendix}
\nocite{*}
%\bibliography{bibfile}

\end{document}